\documentclass[a4paper,10pt,aps,pre,twocolumn,amsmath,amssymb,nofootinbib]{revtex4}
\usepackage{graphicx,color,soul}
\usepackage[colorlinks=true,linkcolor=blue]{hyperref}

\newcommand{\RNum}[1]{\uppercase\expandafter{\romannumeral #1\relax}}

\DeclareMathOperator{\erf}{erf}

\begin{document}

\title{Crystallization of hard spheres revisited. \RNum{1}. Extracting kinetics and free energy landscape from forward flux sampling}

\author{David Richard}
\affiliation{Institut f\"ur Physik, Johannes Gutenberg-Universit\"at Mainz, Staudingerweg 7-9, 55128 Mainz, Germany}
\author{Thomas Speck}
\affiliation{Institut f\"ur Physik, Johannes Gutenberg-Universit\"at Mainz, Staudingerweg 7-9, 55128 Mainz, Germany}

\begin{abstract}
  We investigate the kinetics and the free energy landscape of the crystallization of hard spheres from a supersaturated metastable liquid though direct simulations and forward flux sampling. In this first paper, we describe and test two different ways to reconstruct the free energy barriers from the sampled steady state probability distribution of cluster sizes without sampling the equilibrium distribution. The first method is based on mean first passage times, the second on splitting probabilities. We verify both methods for a single particle moving in a double-well potential. For the nucleation of hard spheres, these methods allow to probe a wide range of supersaturations, and to reconstruct the kinetics and the free energy landscape from the same simulation. Results are consistent with the scaling predicted by classical nucleation theory although a quantitative fit requires a rather large, effective interfacial tension.
\end{abstract}

\maketitle

% ---- introduction ----
\section{Introduction}

Understanding and predicting the kinetics of phase transformations~\cite{kashchiev2000nucleation,kelton2010nucleation} is important in a range of applications from zeolites \cite{lupulescu2014situ} to protein crystallization \cite{sauter2015real} to the correct representation of clouds in climate models \cite{demott2010predicting,hudait2014ice}. After an external change of, say, temperature the current state of the system becomes unstable with respect to the new thermodynamically stable phase. However, due to the free energy cost of an interface between both phases, this transformation does not occur immediately but is driven by rare thermal fluctuations. Atomistic simulations of finite systems are challenged by exponentially small crystallization rates, and a plethora of advanced numerical techniques has been developed such as: umbrella sampling (US)~\cite{auer2001prediction}, metadynamics~\cite{barducci2008well}, transition interface sampling (TIS)~\cite{van2003novel,moroni2005simultaneous}, forward flux sampling (FFS)~\cite{allen2006forward} (and generalizations \cite{berryman2010sampling,becker2012non}), and seeding techniques~\cite{zhang2010weighted,berryman2010sampling,zimmermann2015,statt2015finite,espinosa2016seeding}. These methods were first applied to study simple model systems such as nucleation in the Ising model~\cite{ryu2010numerical} and condensation~\cite{ten1998computer} and crystallization~\cite{trudu2006freezing,jungblut2011crystallization} of Lennard-Jones fluids. More recently, they have been applied to more complex systems and processes such as water evaporation and freezing~\cite{li2011homogeneous,lechner2011role,li2013ice,haji2015direct,menzl2016molecular}, and the crystallization of methane hydrates~\cite{lauricella2014methane,bi2014probing} and Nickel~\cite{bokeloh2011nucleation}.

In this series of papers, we revisit the crystallization of hard spheres in connection with classical nucleation theory (CNT). Hard spheres are very compelling as a testing ground since, conceptually, the model is very simple (only controlled by density) but already exhibits the qualitative behavior found in a wide range of soft-matter systems. Crystallization of hard spheres has been studied extensively in computer simulations~\cite{auer2004numerical,schilling2010precursor,dijkstra10,dijkstra11,russo2012microscopic} and in experiments of colloidal suspensions~\cite{pusey1986phase,harland1997crystallization,iacopini2009crystallization,taffs2013structure,palberg2014crystallization}. While different numerical techniques yield the same nucleation rates, these strongly disagree with experimental results at low supersaturation, which remains an unsolved issue.

Part~\RNum{1} is dedicated to applying and improving a numerical technique to hard spheres that extracts the equilibrium free energy as a function of the order parameter from dynamically biased forward flux sampling. This approach allows to obtain both static and dynamic information (free energy barrier, nucleation rate, collective diffusion coefficient, \emph{etc.}) from the same simulation. Compared to unbiased simulations it allows to study smaller supersaturations with larger nucleation barriers ranging from $10k_BT$ to $50k_BT$. We will discuss two variations based on the sampling of mean first passages times and splitting probabilities in addition to the steady state distribution. In part~\RNum{2}, we will revisit the thermodynamic modeling of the crystallization process on which CNT rests. Based on the nucleation barriers computed from unbiased and FFS simulations, and employing the nucleation theorem~\cite{kashchiev2000nucleation}, we will calculate the pressure and interfacial tension of finite droplets down to small supersaturation with droplets of several thousand particles.

The prevalent view on classical nucleation theory is through the reversible work required to form an ``embryo'' of the novel phase within the (metastable) mother phase. Assuming a spherical shape and, more importantly, that the embryo is characterized by bulk properties leads to a compact expression for the free energy as a function of the embryo size, from the maximum of which one extracts the barrier and the critical nucleus size. Success or failure of CNT is then often judged by how well these final expressions fit the numerical and experimental data. Undetermined factors are subsumed into an ``effective'' interfacial tension, which sacrifices its thermodynamic meaning. This is quite unsatisfactory since such a procedure severely limits the predictive power of CNT.

In the following, we will use the term CNT in this restricted sense in agreement with the current literature. However, the physical picture developed in the founding works by Becker and Doring~\cite{becker1935kinetische} and Zeldovich~\cite{zeldovich1942theory} is much more detailed and not restricted to these additional assumptions. Instead, the two basic requirements are: (i)~the transformation is described by a one-dimensional reaction coordinate through configuration space, and (ii)~the stochastic evolution along this reaction coordinate is slow and Markovian. The dynamics of the system is then described as a master equation or, equivalently, in the continuous limit through a Fokker-Planck equation for the probability to find the system in a microstate compatible with a given value for the reaction coordinate~\cite{peters2016}. While assumption (ii) holds quite generally, (i) has to be checked carefully and many systems indeed follow pathways that require a description with two or more collective variables~\cite{savage2009experimental,sear2012non,sleutel2014role,Dijkstra15,peng2015two}. The term \emph{non-classical} nucleation can then either mean a departure from the additional assumptions for CNT, or, in a stronger sense, a violation of assumption (i).

Numerical methods like umbrella sampling and metadynamics allow to reconstruct the equilibrium free energy as a function of the nucleus size and thus yield the reversible work for the phase transformation. These methods employ constrained dynamics through biasing the potential and forces. In contrast, TIS and FFS are based on the unbiased microscopic dynamics with access to rare events through a branching of trajectories. Exploiting detailed balance allows to extract the free energy landscape (in addition to the nucleation rate), but one needs to sample both forward (growth) and the backward (melting) trajectories~\cite{valeriani2007computing}, which is computationally expensive. Wedekind and Reguera have described a way to reconstruct the equilibrium free energy landscape using only forward trajectories through computing the forward steady state probability of cluster sizes and the mean first passage time (MFPT) to reach a cluster of a given size~\cite{wedekind2008kinetic}. This approach, referred to as MFPT-FFS, was applied for the condensation and crystallization of a Lennard-Jones fluid in brute-force simulations~\cite{wedekind2009crossover,lundrigan2009test}. Recently, this method was extended in combination with FFS to study the nucleation in the Ising model for barriers not accessible via direct brute force runs~\cite{thapar2015simultaneous}.

An alternative approach that does not require to compute the MFPT is based on splitting probabilities, which have been used mainly in the context of folding and unfolding of biomolecules~\cite{chodera2011,manuel2015}. The empirical splitting probability along the chosen order parameter is directly computed in FFS, and thus offers a direct way to extract barriers. We first test this method for a Brownian particle trapped in a double-well potential using a combination of brute force simulations and FFS, and provide a comparison with the result given by MFPT-FFS. We then revisit the arguably best studied off-lattice model system: hard spheres.

% ---- methods ----
\section{Methods}

\subsection{Formalism}
\label{sec:form}

\begin{figure}[b!]
  \centering
  \includegraphics[width=.9\linewidth]{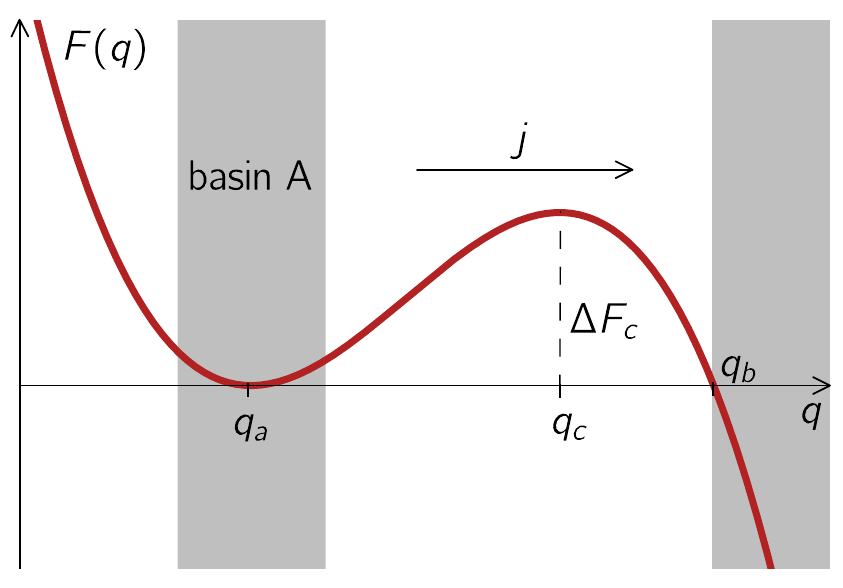}
  \caption{\textbf{Metastable state.}~Sketch of a typical free energy landscape along a single reaction coordinate $q$. The system is trapped in basin $A$ and has to overcome a (large) barrier $\Delta F$ at $q_c$. An absorbing boundary is placed at $q_b$ and a reflecting boundary at $-\infty$ via a steep potential, creating a constant current $j$ of probability flowing in the direction of positive $q$ and a steady probability distribution $P^+(q)$.}
  \label{fig:problem}
\end{figure}

Since the relevant expressions are scattered through the literature, we start by briefly outlining the formalism used in our simulations to reconstruct the free energy landscape from the non-equilibrium steady state sampled by techniques such as FFS. We consider rare transitions that are well captured by a single reaction coordinate $q$ with restricted free energy $F(q)$ as depicted in Fig.~\ref{fig:problem}. The system is originally located in a deep basin $A$ around $q_a$. A rare fluctuation is needed to overcome a large barrier $\Delta F_c=F(q_c)-F(q_a)$ located at $q_c$. The time evolution of the probability distribution $P(q,t)$ to find the system at $q$ at time $t$ is given by the Fokker-Planck (FP) equation~\cite{gardiner1985stochastic}
\begin{equation}
\label{eq:FP}
\frac{\partial P(q,t)}{\partial t}  =  \frac{\partial}{\partial q} D(q)\Bigg[\frac{\partial \beta F(q)}{\partial q}+\frac{\partial}{\partial q}\Bigg] P(q,t)= -\frac{\partial j(q,t)}{\partial q},
\end{equation}
where $j(q,t)$ is the probability current and $D(q)$ the diffusion coefficient of the collective variable $q$. In a steady state, $\partial P(q,t)/\partial t=0$ and $j$ is time independent and constant for all values of $q$. It is convenient to express the current as
\begin{equation}
\label{eq:current}
j(q,t)=-D(q)e^{-\beta F(q)}\frac{\partial}{\partial q}P(q,t)e^{\beta F(q)}.
\end{equation}
In equilibrium, the current vanishes ($j=0$) and the solution of the Fokker-Planck equation yields the Boltzmann factor $P(q)\propto e^{-\beta F(q)}$ linking it to the free energy $F(q)$.

We are interested in the rate to leave basin $A$, which can be expressed as a mean first passage time (MFPT)
\begin{equation}
  \label{eq:mfpt}
  \tau^+(q) = \int_{q_a}^{q}dy\;\frac{e^{\beta F(y)}}{D(y)}\int_{-\infty}^{y}dz\; 
  e^{-\beta F(z)},
\end{equation}
to reach any coordinate $q$ from basin $A$. This is an exact result that depends on the full functional form for the free energy. To obtain a more compact and useful expression, we introduce an absorbing boundary at $q_b>q_c$ sufficiently far from the barrier. Demanding conservation of probability now implies a non-vanishing current $j$ (absorption at $q_b$ is compensated by injection at $q\ll q_a$) and thus a \emph{non-equilibrium} steady state with solution $P^+(q)\neq P(q)$ of Eq.~(\ref{eq:FP}). Note that $P^+(q_b)=0$. Expanding the free energy $F(q)\approx F(q_c)-(1/2)|F''(q_c)|(q-q_c)^2$ allows a saddle-point integration of Eq.~(\ref{eq:current}) with the resulting Kramer's rate~\cite{kramers1940brownian}
\begin{equation}
  \label{eq:kramer}
  j\approx D(q_c)Z(q_c)\frac{e^{-\beta\Delta F_c}}{\int_{A} dx\; e^{-\beta F(x)}},
\end{equation}
and of the MFPT [Eq.~(\ref{eq:mfpt})] with result
\begin{equation}
  \label{eq:mfpt_th}
  \tau^+(q)\approx\frac{1}{2j}\{1+\erf[\sqrt{\pi}Z_c(q-q_c)]\}.
\end{equation}
Here, $Z_c=Z(q_c)=\sqrt{\frac{|F''(q_c)|}{2\pi k_BT}}$ is called the Zeldovich factor related to the width of the barrier. As expected for an activated process, we see that the crossing rate will be dominated by the Boltzmann weight of the potential barrier $e^{-\beta\Delta F_c}$. One easily finds $\tau^+(q_b)\simeq 1/j$, which shows that the current corresponds to the escape rate from basin $A$ as expected. Eq.~(\ref{eq:mfpt_th}) has been used to extract the escape rate $j$, the critical barrier position $q_c$, and the Zeldovich factor $Z_c$ for the gas-liquid nucleation of a Lennard-Jones fluid~\cite{wedekind2007new}.

From Eq.~(\ref{eq:current}) it is also possible to link the underlying free energy $F(q)$ to the forward probability distribution $P^+(q)$ through
\begin{equation}
\frac{\partial \beta F(q)}{\partial q} = -\frac{\partial \ln P^+(q)}{\partial q}-\frac{j}{D(q)P^+(q)}.
\end{equation}
Through integration one can compute the free energy difference between two reference state points. Unfortunately, in practice one needs to know the diffusion coefficient $D(q)$, which is expensive to compute. To circumvent this issue, Wedekind and Reguera have demonstrated that the steady state probability $P^+(q)$ can be combined with the MFPT $\tau^+(q)$ to determine
\begin{equation}
  \label{eq:v_mfpt}
  \beta F(q) = \beta F(q_0) +  \ln\frac{B(q)}{B(q_0)} 
  - \int_{q_0}^q dy\;\frac{1}{B(y)}
\end{equation}
with
\begin{equation}
  B(q) = -\frac{1}{P^+(q)}\Bigg[\int_q^{q_b} dy\; 
  P^+(y)-\frac{\tau^+(q_b)-\tau^+(q)}{\tau^+(q_b)}\Bigg],
\end{equation}
where $q_0$ is an arbitrary lower integration bound~\cite{wedekind2008kinetic}. By definition $B(q)=D(q)/[\partial\tau^+(q)/\partial q]$, which can then be used to determine $D(q)$~\cite{wedekind2008kinetic,lundrigan2009test}. This method was employed to study the gas-liquid~\cite{wedekind2009crossover} and liquid-solid~\cite{lundrigan2009test} nucleation in a Lennard-Jones fluid.

An alternative route to the free energy that does not involve the MFPT $\tau^+(q)$ is based on splitting probabilities. Employing a discrete master equation~\cite{abraham1969multistate,ter2003determination,karlin2014first}, one finds
\begin{equation}
  \label{eq:pst_peq}
  P^+(q) = P(q)P_A(q)
\end{equation}
linking the steady state probability distribution $P^+(q)$ to the equilibrium distribution $P(q)$ (appendix~\ref{ap:pst_peq}). They are related through the splitting probability
\begin{equation}
  \label{eq:pa}
  P_A(q) = \int_q^{q_b} \frac{dx}{f(x)P(x)} \Bigg(\int_{q_A}^{q_b} 
  \frac{dx}{f(x)P(x)}\Bigg)^{-1}
\end{equation}
to fall back to basin $A$ from coordinate $q$, where $f(q)$ are attachment rates (appendix~\ref{ap:pa_pb}). The probability to reach basin $B$ is $P_B(q)=1-P_A(q)$. One possibility to extract the free energy is to take the derivative of Eq.~(\ref{eq:pa}) with respect to $q$~\cite{manuel2015}, which requires a numerical derivative and the knowledge of $f(q)$. Noticing that $P(q)$ in Eq.~(\ref{eq:pst_peq}) is simply $\propto e^{-\beta F(q)}$, one can derive an expression valid for large barriers using the same steepest descent method as before, yielding
\begin{equation}
  \label{eq:pb_th}
  P_B(q) \approx \frac{1}{2}(1+\erf[\sqrt{\pi}Z_c(q-q_c)])
\end{equation}
independent of the attachment rates $f(q)$. From Eq.~(\ref{eq:pb_th}), the equilibrium probability can then be extracted through
\begin{equation}
  \label{eq:v_pb}
  P(q) = P^+(q)/[1-P_B(q)]
\end{equation}
making use of the forward probability $P^+(q)$. This expression can be used to compute $Z_c$ and $q_c$, where the top of the free energy barrier with $P_B=1/2$ defines the transition state ensemble.

\begin{figure*}[t]
  \includegraphics[scale=1.0]{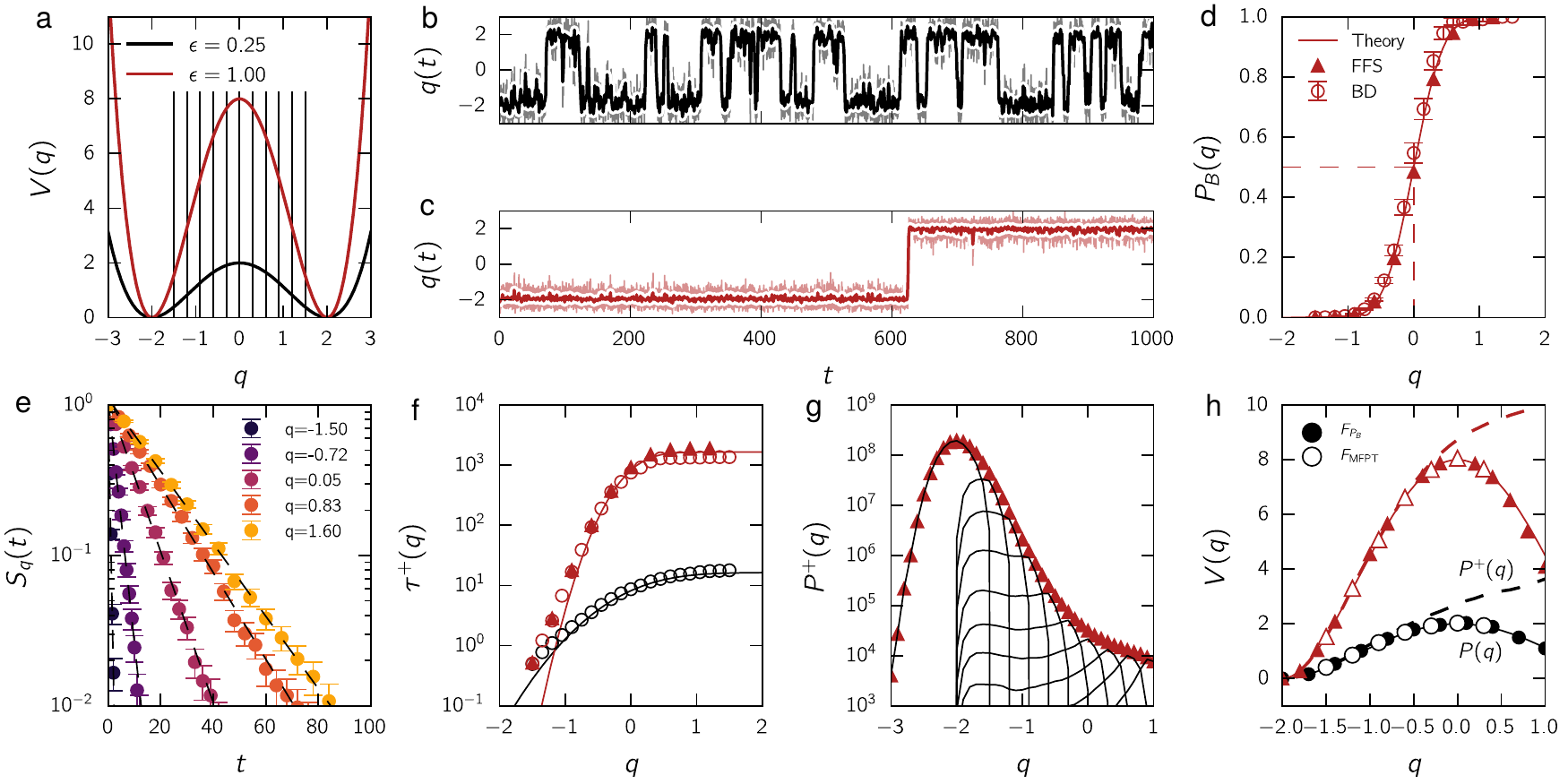}
  \caption{\textbf{Brownian particle in a double-well potential.} (a)~The double-well potential $V(q)$ from Eq.~(\ref{eq:pot}) as function of the particle position $q$ for $\epsilon=1/4$ and $\epsilon=1$. (b,c)~Time evolution of the particle position $q(t)$ and steady state probability $P^+(q)$ for (b) $\epsilon=1/4$ and (c) $\epsilon=1$. (d)~Splitting probability $P_B(q)$ for $\epsilon=1$. The solid line is the analytical solution from Eq.~(\ref{eq:pb_th}). Filled triangles correspond to FFS and open circles to direct Brownian dynamics simulations. (e)~Evolution of the survival probability of \emph{not} reaching the position $q$ within time $t$. Solid lines are linear fits to Eq.~(\ref{eq:survival}). (f)~Evolution of the mean first passage time $\tau^+(q)$ for $\epsilon=1/4$ (black symbols) and $\epsilon=1$ (red symbols). The solid lines are the analytical solution from Eq.~(\ref{eq:kramer}) and Eq.~(\ref{eq:mfpt_th}). (g)~Reconstruction of the steady state probability distribution $P^+(q)$ from FFS using Eq.~(\ref{eq:pst_ffs}). (h)~Reconstruction of the negative logarithm of the equilibrium probability $-\ln P(q)$ based on the splitting probability (filled symbols) and MFPT (open symbols). Also shown are the true potential $V(q)$ (solid lines) and the negative logarithm of the steady state probability distribution $-\ln P^+(q)$ (dashed lines).}
  \label{fig:dbwell}
\end{figure*}

\subsection{Forward flux sampling}
\label{sec:ffs}

Forward flux sampling (FFS) guides the reaction through defining a set of $M$ interfaces at values $q_a<q_i<q_b$ of the reaction coordinate~\cite{allen2006forward}, see the sketch in Fig.~\ref{fig:dbwell}(a). We first compute the MFPT $\tau^+(q_1)$ to reach the first interface from basin $A$. To this end, we collect various configurations at $q=q_1$. In practice $q_1$ should not be chosen too close to $q_A$ to let the system relax and populate the basin $A$ before reaching $q_1$. We then run multiple trajectories from saved configurations to reach either the next interface or to fall back to basin $A$. Their ratio provides the forward transfer probability $P_i=P(q_{i+1}|q_i)$ to reach the next interface. This procedure is continued until we reach $q_b$, whereby we assume that $P_M=P(q_b|q_M)\simeq 1$, \emph{i.e.}, passing the last interface is an (almost) irreversible transition. It should be clear that this procedure samples exactly the non-equilibrium steady state discussed in the previous section~\ref{sec:form}.

From these simulations, we calculate the empirical splitting probability
\begin{equation}
  \label{eq:pb_sim}
  \hat P_B(q_i) = \prod_{k=i}^MP_k
\end{equation}
to reach the absorbing boundary (and thus basin $B$) as an average over trajectories assuming that the transfer probabilities $P_i$ are independent. We evaluate errorbars using standard propagation of errors, see Ref.~\cite{allen2006forward} for details. Combining Eq.~(\ref{eq:mfpt_th}) with Eq.~(\ref{eq:pb_th}) yields $\tau^+(q)=P_B(q)/j$ and thus
\begin{equation}
  \label{eq:k_ffs}
  j = \frac{P_B(q_1)}{\tau^+(q_1)}
\end{equation}
for the reaction rate.

We evaluate the steady state distribution $P^+(q)$ following Ref.~\cite{valeriani2007computing,thapar2015simultaneous}. The probability $P^+(q)$ is computed from gathering all histograms $\pi_i(q)$ from trajectories fired at each interface placed at $q_i$. Since we already know the forward probability to reach a given interface, we can reweigh all $\pi_i(q)$ accordingly, which yields
\begin{equation}
\label{eq:pst_ffs}
KP^+(q) = \pi_{A}(q) + \pi_1(q) +
\sum_{i=2}^{M-1} \pi_i(q) \prod_{k=1}^{i-1}P_k
\end{equation}
with normalization constant $K$, $\pi_{A}(q)$ the distribution of states along the trajectories that started in $A$ ($q<q_A$) and stopped at $q_1$, and $\pi_{q_i}(q)$ the distribution of trajectories that started at $q_i$ and stopped either at $q_{i+1}$ or $q_A$.

At this point there are two alternatives to extract the free energy $F(q)$. For the first method we need to compute $\tau^+(q)$. To determine the MFPT at every interface we follow Ref.~\cite{thapar2015simultaneous}, where the authors derived an expression for $\tau^+(q)$ based on the crossing probabilities,
\begin{equation}
  \label{eq:mfpt_ffs}
  \tau^+(q_{i+1}) = \frac{\tau^+(q_i)+\tau^-(q_i)}{P_i}
  + \tau^+(q_{i+1}|q_i)-\tau^-(q_i).
\end{equation}
Here, $\tau^+(q_i)$ and $\tau^-(q_i)$ are the MFPT to reach $q_i$ from $A$ and to fall back from $q_i$ to $A$, respectively, and $\tau^+(q_{i+1}|q_i)$ is the MFPT to go from $q_i$ to $q_{i+1}$. We can recursively evaluate $\tau^+(q_{i+1})$ if we start in a basin $A$ and evaluate $\tau^+(q_1)$ and $\tau^-(q_1)$ from our brute force runs. Note that we have already evaluated $\tau^+(q_1)$ for computing $j$. One can then combine $P^+(q)$ and $\tau^+(q)$ together with Eq.~(\ref{eq:v_mfpt}) to extract $F(q)$. The second method is to directly use the already computed $\hat P_B(q)$ from Eq.~(\ref{eq:pb_sim}) together with $P^+(q)$ to compute the equilibrium distribution $P(q)$ via Eq.~(\ref{eq:v_pb}), leading to $F(q)$.

\subsection{Brute force}
\label{sec:md}

For small to intermediate barriers, it is sufficient to harvest trajectories from direct simulations. By initializing the system in basin $A$ and terminating once we have reached basin $B$, we again sample a steady state instead of equilibrium (which would require to sample many barrier crossing). We evaluate the MFPT $\tau^+(q)$ through computing, from independent runs, the survival probability $S_q(t)$ that the system at time $t$ did not leave a domain bounded by $q>q_a$. For a Markov process, this survival probability decays exponentially,
\begin{equation}
  \label{eq:survival}
  S_q(t) = e^{-(t-t_0)/\tau^+(q)}.
\end{equation}
Following Ref.~\cite{chkonia2009evaluating}, we choose to include an offset $t_0$ to take into account the initial transient regime to sample basin $A$. From this method, we extract the evolution of $\tau^+(q)$, which includes the reaction rate $j=1/\tau^+(q_b)$. We evaluate the splitting probability $\hat P_B(q)$ as in FFS by placing fictive interfaces at various values $q_i$. From all independent runs and recorded histories of the system being at a given reaction coordinate $q$, one can evaluate accurately all transfer probabilities $P_i$ by counting when the system was crossing a fictive interface at $q_i$ and reaching the next one at $q_{i+1}>q_i$ before falling back to basin $A$. The final part of the analysis consists of building the overall histogram of the visited values of $q$ to construct the normalized probability distribution $P^+(q)$.

\subsection{Trapped particle in a double-well potential}

We now illustrate the formalism for a simple model system: a single particle moving in a one-dimensional potential. In this case, the expressions derived in Sec.~\ref{sec:form} are exact and the reaction coordinate corresponds to the position of the particle and the free energy to the potential energy. We choose the double well potential
\begin{equation}
  \label{eq:pot}
  V(q) = \epsilon q^2\left(\frac{1}{2}q^2-4\right),
\end{equation}
where $\epsilon$ controls the height of the barrier, see Fig.~\ref{fig:dbwell}(a) for $\epsilon=1/4$ and $\epsilon=1$. We indicate the two local basins A and B placed at $q=-2$ and $q=2$, respectively. The barrier position is located at $q_c=0$. To integrate the particle motion, we employ Brownian Dynamics (BD) with
\begin{equation}
\label{eq:bd}
q(t+\Delta t) = q(t) - \frac{\Delta t}{k_BT}V'(q(t)) + \sqrt{2\Delta t}\zeta_i,
\end{equation}
where the $\zeta_i$ are Gaussian noises with zero mean and unit variance and $\Delta t$ is the time step. As the unit of time, we employ the Brownian time $\tau_B=l^2/D_0$ with constant translational diffusion coefficient $D_0$ and length scale $l$. The time step is set to $\Delta t = 10^{-5}\tau_B$. The FFS interfaces are indicated by vertical lines in Fig.~\ref{fig:dbwell}(a). In Fig.~\ref{fig:dbwell}(b,c), the typical time evolution of the particle position for $1000 \tau_B$ is depicted for the two different values of $\epsilon$ with $\epsilon=1/4$ and $\epsilon=1$, respectively. As expected for the lowest barrier ($2k_BT$), the particle transitions more often between the two basins. For the larger barrier ($8k_BT$), we observe during the same observation time only one crossing event with a long residence time in basin $A$, which is characteristic for a metastable state.

In Fig.~\ref{fig:dbwell}(d), we show the evolution of the splitting probability $P_B(q)$ as a function of the particle position. Note that we only show the result for $\epsilon=1$ for the sake of visibility. We find a very good agreement between the simulation results based on Eq.~\ref{eq:pb_sim} and the analytical solution based on Eq.~\ref{eq:pb_th}, where $q_c=0$ and $Z_c=2\sqrt{\epsilon/\pi}$. We then present in Fig.~\ref{fig:dbwell}(e) the survival probability $S_q(t)$ for $\epsilon=1/4$ and various values of $q$. The dashed line corresponds to the linear fit used to extract $\tau^+(q)$ according to Eq.~\ref{eq:survival}. We show in Fig.~\ref{fig:dbwell}(f) the evolution of the MFPT for $256$ independent runs of $1000\tau_B$. We find a good agreement between the computed MFPT and the analytical Eq.~\ref{eq:pb_th}, where we use the Kramer approximation for the rate $j$. Note that for $\epsilon=1$ only $20\%$ of the runs show a crossing event but we can already estimate the correct escape rate with only minor deviation from the prediction. To recover $j$ with higher accuracy, we employ FFS-MFPT with Eq.~\ref{eq:mfpt_ffs}. We find a very good agreement with the MD results for small $q$ and the MFPT converges, after the barrier, to the correct inverse rate $1/j$. We additionally show in Fig.~\ref{fig:dbwell}(g), the resulting FFS reconstruction of the steady state distribution $P^+(q)$ before normalization using Eq.~\ref{eq:pst_ffs}. This analysis leads to the reconstruction of the underlying potential using the splitting probability and MFPT, see Fig.~\ref{fig:dbwell}(h). We indicate in solid and dashed lines the potential $V(q)$ and the negative logarithm of the steady state probability $-\ln[P^+(q)]$. We see that the steady state probability gives a correct description of the potential within basin $A$ but does not capture the barrier and instead continues to increase. The free energy reconstruction $F_{P_B}$ from $P_B(q)$ using Eq.~\ref{eq:v_mfpt} is shown in colored filled symbols and the reconstruction $F_{\text{MFPT}}$ from $\tau^+(q)$ with Eq.~\ref{eq:v_mfpt} is shown in white filled symbols. We find that both methods  correctly reproduce the potential and thus give access to the barrier. The second method has the advantage that in practice one can extract the rate and the free energy profile using only the transfer probabilities without the need to compute the MFPT.

% ---- hard spheres ----
\section{Crystallization of hard spheres}

\subsection{Model and mapping}

We now turn to hard spheres. In order to solve the equations of motion for discontinuous hard potentials there are basically two options: event-driven molecular dynamics~\cite{rapa80} or standard molecular dynamics (MD) with continuous forces, the results of which are then mapped onto hard spheres. Here we follow the latter, which is an established technique and has been shown to quantitatively reproduce the static and dynamic behavior of hard spheres (for a discussion on experiments with colloidal particles, see Ref.~\cite{royall2013}). Specifically, we simulate the crystallization of a liquid composed of $N=5,000$ monodisperse particles interacting via the Weeks-Chandler-Andersen (WCA) potential
\begin{equation}
  \label{eq:wcapot}
  u_\text{WCA}(r) = 4\epsilon [(\sigma/r)^{12}-(\sigma/r)^{6}+1/4 ]
\end{equation}
for $r<2^{1/6}\sigma$ and zero otherwise. We work at constant volume $V$ and temperature $T$. The dynamics is coupled with the stochastic Lowe-Anderson thermostat~\cite{Lowe06}, which applies binary collisions for pairs of particle within a collision radius $R_T$. The collision rule conserves both linear and angular momentum. It is characterized by the bath collision frequency $\Gamma=n_\text{col}/(N\Delta t)$ with $n_\text{col}$ the number of collisions during the time step $\Delta t$. For low frequency of collision $\Gamma < 100$, the thermostat does not impact the diffusion coefficient of the liquid~\cite{Lowe06}. We use $\Gamma=10$ for all simulations.

\begin{figure}[t]
  \includegraphics[scale=1.0]{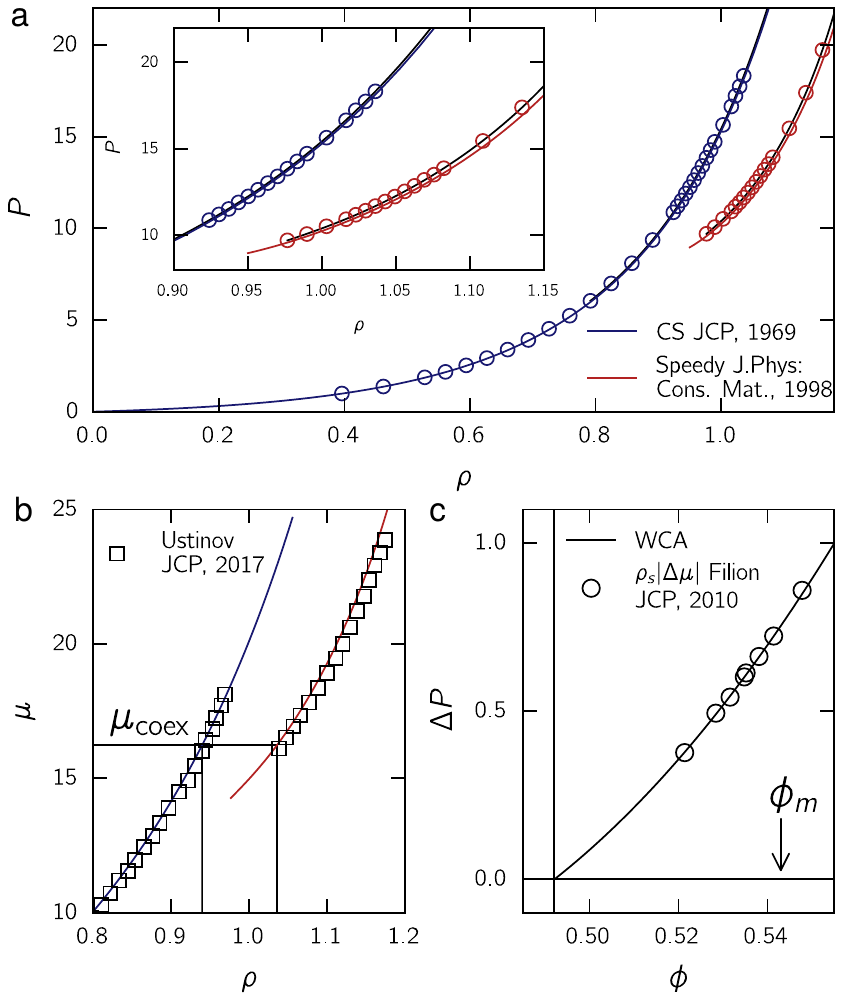}
  \caption{\textbf{Mapping WCA onto hard spheres.}~(a) Pressure against the density for the liquid (blue) and solid (red) branches. Empty circles are MD data of the WCA fluid ($\beta\epsilon=40$). Black solid lines are polynomial fits of the data. Blue~\cite{Carnahan69} and red~\cite{speedy1998pressure} solid lines are literature equation of state for hard spheres. (b)~Chemical potential as a function of the density in hard-sphere units. Blue and red lines are WCA data and empty squares are the hard sphere data from Ref.~\cite{ustinov2017thermodynamics}. The two black vertical lines corresponds to the coexisting densities of the WCA fluid from a common tangent construction. (c)~Evolution of the pressure difference between the liquid and solid phase at the same ambient chemical potential as a function of the packing fraction. The solid line corresponds to the WCA fluid and empty circles are $\Delta P=\rho_s|\mu_s-\mu_l|$ from the hard sphere data from Ref.~\cite{dijkstra10}.}
  \label{fig:mapping}
\end{figure}

The mapping consists of assigning an effective diameter $d$ to the WCA particles that corresponds to the diameter of true hard spheres. Here we follow the procedure of Filion \emph{et al.}~\cite{dijkstra11} and map the freezing density $\rho_f\simeq0.712\sigma^{-3}$ of WCA to the freezing packing fraction $\phi_f\simeq0.492$ of hard spheres. With $\phi=(\pi/6)\rho d^3$ we thus obtain $d\simeq1.097\sigma$. Note that this approach gives similar results compared to the Barker-Henderson approximation~\cite{Barker70} and WCA perturbation theory~\cite{Chandler71} (less than 1\% of difference). From now on we employ dimensionless quantities with $d$ the unit of length and (free) energies given in units of the thermal energy, $k_BT$. The strength of the WCA potential is set to $\epsilon=40k_BT$. To map the time, we follow Ref.~\cite{Noyola13} and scale the MD time via an ``atomistic'' Brownian time as $t^*=t/\tau_B$, with $\tau_B=d^2/D_0$ where $D_0=\frac{3}{8}\sqrt{\frac{k_BT}{\pi m}}\frac{1}{\rho d^2}$ is the short-time self-diffusion coefficient obtained from the Chapman-Enskog kinetic theory of gases~\cite{Chapman70}. This procedure is known to accurately map the MD relaxation time onto BD for relatively dense system ($\phi > \phi_f$)~\cite{Noyola13}. Previously, we have tested this mapping by comparing crystallization rates to BD results for the same system~\cite{richard2015role}.

To validate the mapping of our WCA fluid onto hard spheres, we present in Fig.~\ref{fig:mapping}(a) the pressure as a function of the density. We compare our MD data with literature equations of state for the bulk fluid [$P_l(\rho)$] and solid [fcc crystal, $P_s(\rho)$]. We find a very good agreement of the liquid branch with Carnahan-Starling~\cite{Carnahan69} and of the solid branch with Speedy~\cite{speedy1998pressure}. There are some small discrepancies for the solid branch, which are highlighted in the inset. To compute the chemical potential vs. the density, we perform a thermodynamic integration with the reference state points taken from Ref.~\cite{dijkstra11}. This procedure yields the functions $\mu_l(\rho)$ and $\mu_s(\rho)$, which we compare in Fig.~\ref{fig:mapping}(b) with recent numerical results from Ref.~\cite{ustinov2017thermodynamics}. As for the pressure, we find a very good agreement between WCA and hard sphere data with a slight shift of the WCA solid branch towards lower densities. The small shift in pressure and chemical potential does not have any impact on $\Delta P=P_s-P_l$ (at the same ambient chemical potential) as seen in Fig.~\ref{fig:mapping}(c) when plotted as a function of the packing fraction $\phi$.

\subsection{Collective variable}
\label{sec:cv}

Numerous numerical investigations~\cite{auer2001prediction,dijkstra10,dijkstra11,russo2012microscopic,richard2015role} have established that the nucleation of hard spheres is rather well described by a single collective variable related to the size of solid droplets (although adding structural information might improve the reaction coordinate as shown for Lennard-Jones~\cite{beckham2011,jungblut2013}). To construct this collective variable from particle configurations, we apply the same method as in Ref.~\cite{dijkstra11}. The local bond-orientational parameter~\cite{Steinhardt1983,Dell08}
\begin{equation}
  \label{eq:order}
  q_{l,m}(i) = \frac{1}{N_n(i)}\sum_{j=1}^{N_n(i)} Y_{l,m}(\theta_{i,j},\varphi_{i,j})
\end{equation}
is evaluated for particle $i$, where $Y_{l,m}(\theta,\varphi)$ are spherical harmonics and $N_n$ is the number of neighbors within distance $r_{ij} < 1.5\sigma$. We construct a bond network through the scalar product
\begin{equation}
  \label{eq:scalar}
  d(i,j) = \frac{\sum_{m=-l}^{l} q_{l,m}(i) q_{l,m}^\ast(j)}{(\sum_{m=-l}^{l} |q_{l,m}(i)|^2)^{1/2} (\sum_{m=-l}^{l} |q_{l,m}(j)|^2)^{1/2}}
\end{equation}
using $l=6$ with $d(i,j) > 0.7$ defining a bond. A particle is defined as "solid-like" if the number of bonds $\xi\ge 9$, and clusters are constructed from mutually bonded solid-like particles.

The method discussed above computes the free energy $F_{\text{big}}(n)$ of the \emph{largest} cluster in the system. It is important to point out that $F_{\text{big}}(n)$ does not directly correspond to the actual free energy $F(n)$ associated with the formation of any cluster of size $n$, which is the quantity of interest for CNT~\cite{reiss1999some,auer2004numerical}. The function $F(n)$ increases monotonically with $n$ and is invariant with respect to the system size, which is not the case for $F_{\text{big}}(n)$~\cite{maibaum2008comment}. To circumvent this issue and reconstruct $F(n)$ from $F_{\text{big}}(n)$, we have employed the same method as in Refs.~\cite{lundrigan2009test,leitold2016nucleation}. Further details are provided in appendix~\ref{ap:fe_reconstruct}.

\subsection{Results}

\begin{figure*}[t]
  \includegraphics[scale=1.0]{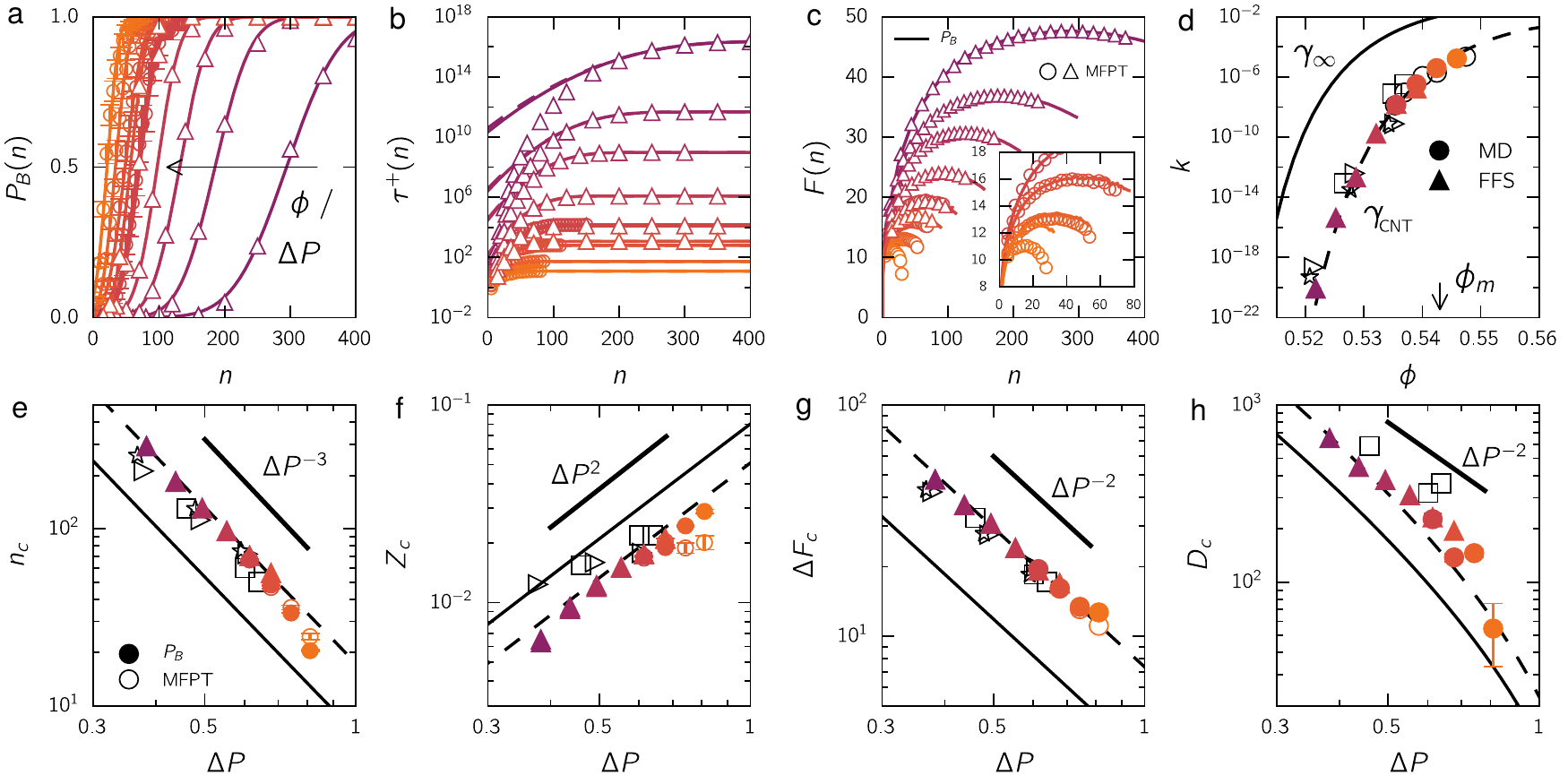}
  \caption{\textbf{Crystallization of hard spheres.} (a)~Splitting probability $\hat P_B(n)$ as function of the largest nucleus size $n$. Solid lines are fits to Eq.~\ref{eq:pb_th}. The arrow indicates increasing packing fraction $\phi$ and pressure difference $\Delta P$. (b)~MFPT $\tau^+(n)$ as function the largest nucleus size $n$ (symbols). Solid lines are fits to Eq.~\ref{eq:mfpt_th}. (c)~Reconstruction of the free energy landscape from Eq.~\ref{eq:pst_peq} and Eq.~\ref{eq:v_mfpt}. Solid lines are $F_{P_B}(n)$, circles and triangles corresponds to $F_{\text{MFPT}}(n)$. The inset show a zoom of $\Delta F(n)$ for large packing fractions. (d) Nucleation rate $k=j/V$ as function of the packing fraction $\phi$. (e-h) Evolution of (e) the critical nucleus size $n_c$, (f) the Zeldovich factor $Z_c$, (g) the nucleation barrier $\Delta F_c$ and (h) the critical diffusion coefficient $D_c$ as functions of the pressure difference $\Delta P$. In all panels, discs correspond to unbiased and triangles to FFS simulations. In (e-h), open symbols are values extracted through MFPT and closed symbols from the splitting probability $P_B$. Thick straight lines indicate the slopes of the asymptotic power laws for $\Delta P\to0$. Black lines correspond to the CNT expression with bulk interfacial tension $\gamma_{\infty}\simeq0.56$ (solid) and effective interfacial tension $\gamma_{\text{CNT}}=0.76$ (dashed). The different colors from orange to purple are various packing fractions ranging from $\phi\simeq0.523$ to $\phi\simeq0.547$. Additional data points are taken from: ($\star$)~Ref.~\cite{auer2004numerical}, ($\square$)~Ref.~\cite{dijkstra11}, and ($\triangleright$)~Ref.~\cite{dijkstra10}.}
  \label{fig:hs}
\end{figure*}

We now present results for both ``brute force'' and FFS simulations. Random configurations of dense hard spheres without initial solid clusters were created with the algorithm developed by Clarke and Wiley~\cite{Wiley87}, where the non-overlapping distance between particles was chosen to be equal to the effective diameter $d_\text{eff}$. For the direct simulations, we have harvested 128 independent crystallization trajectories with the cluster size $n$ recorded every $\Delta t \simeq 0.006 \tau_B$. For FFS, we collect 100 configurations at the first interface placed at $n=15$ for $\phi>0.528$ and at $n=12$ for the lower packing fractions. The number of trials to reach the next interface is typically between $200-10000$, where the lowest value is used at the end of the crystallization process when the transfer probability has reached $P_{q_i}>0.9$. Interfaces are placed so that $P_{q_i}>0.001$ and $\Delta n=n_{i+1}-n_i>10$. We have checked that errors for $P_B(q)$ are smaller than symbol sizes. We have recorded $n$ every $\Delta t \simeq 0.003 \tau_B$.

We apply the same method described previously to study the crystallization of hard spheres for various packing fractions ranging from $\phi\simeq0.522$, which is close to the middle of the coexistence region, to $\phi\simeq0.547$, which is above the melting point $\phi_m\simeq0.543$. In Fig.~\ref{fig:hs}(a), we first discuss the change of the splitting probability $P_B(n)$ when we progressively decrease the supersaturation with decreasing pressure difference $\Delta P(\phi)=P_s-P_l$ between solid (pressure $P_s$ at density $\rho_s$) and liquid ($P_l$ at $\rho_l=6\phi/\pi$) at the same ambient chemical potential, which is the thermodynamic driving force for the nucleation process. For the largest packing fraction $\phi\simeq0.547$, we observe a quick increase of $P_B(n)$ for small cluster sizes, from which we estimate the critical nucleus size at $P_B=1/2$ to be approx. $20$ particles. This rather small critical size is not surprising since we already passed through the melting point and thus we are dealing with a quick nucleation process, where no clear distinction between the induction and growth regime can be made anymore. This observation is confirmed by the small nucleation barrier. Decreasing $\phi$, we observe the same qualitative behavior for $P_B(n)$ at all packing fractions, but the critical nucleus size $n_c$ defined through $P(n_c)=1/2$ moves to larger sizes.

In Fig.~\ref{fig:hs}(b), the evolution of the MFPT shows the same trend. The transition at which the MFPT saturates moves to larger cluster sizes when decreasing the packing fraction $\phi$. Additionally, we see that the position of the plateau for $\tau^+(n)$, which corresponds to the inverse escape rate $1/j$ (nucleation time), is increasing logarithmically, which is what we expect from an activated process, cf. Eq.~\ref{eq:kramer}. From this data, we extract $j$ when reaching $P_B(n)\simeq1$. We compare the fit of $\tau^+(n)$ (solid lines) and the resulting fit of $P_B(n)$ from Fig.~\ref{fig:hs}(a) multiplied by $1/j$ (dashed lines). We find a very good agreement, which justifies the use of Eq.~\ref{eq:pb_th} and Eq.~\ref{eq:mfpt_th}.

Having extracted $n_c$ and $Z_c$ from these fits, we can proceed to reconstruct the free energy landscape. In Fig.~\ref{fig:hs}(c), we present a comparison between $\Delta F_{P_B}(n)$ and $\Delta F_{\text{MFPT}}(n)$. We find a very good agreement between both methods with some small discrepancies at higher packing fractions, see inset of Fig.~\ref{fig:hs}(c). In this regime, the nucleation process is somewhat instantaneous and thus the evaluation of $P^+$ and $\tau^+$ becomes more difficult and will additionally be sensitive to the way the system is prepared in the initial supersaturated state. Additionally, in this regime we expect that all analytical results based on expanding the free energy at the top of the barrier will start to fail. Finally, it is worth noticing that beyond the melting point, multiple droplets might merge together (\emph{e.g.}, as shown for the gas-liquid nucleation \cite{yasuoka1998molecular}). Thus, the assumption that the process is a first-order reaction, where droplets are independent, will also break down. In contrast, for large barriers the agreement is perfect, and we find for the lowest packing fraction $\phi\simeq0.522$ a nucleation barrier of $\Delta F_c\simeq 48k_BT$. This result agrees with previous results using umbrella sampling~\cite{auer2004numerical,auer2001prediction,dijkstra11}.

As a remark we note that during the computation of $\Delta F_{\text{MFPT}}(n)$, we evaluate $B(n)$ and $\partial\tau^+(n)/\partial n$, which allows the determination of $D_c$ through $B(n_c)/(\partial\tau^+(n)/\partial n)|_{n=n_c}$. Despite the fact that this is a formal relation, in practice it means to compute, on top of the free energy barrier, the ratio between two very large numbers, which implies large numerical errors as demonstrated in Ref.~\cite{lundrigan2009test} for the crystallization of a Lennard-Jones liquid. We decided to compute $D_c$ as the average of $D(n)$ over the range $0.8n_c<n<1.2n_c$ and to take the standard deviation as an estimation for the error.

In Fig.~\ref{fig:hs}(d), we plot the evolution of the crystallization rate $k=j/V$ \emph{vs.} the packing fraction $\phi$. We compare our results with previous numerical studies of exact hard spheres~\cite{auer2004numerical,dijkstra10} and the WCA fluid mapped onto hard spheres~\cite{dijkstra11}. We find an excellent agreement between our results and earlier studies. The nucleation rate spans approx. $15$ orders of magnitude between the middle of the coexisting region and the melting point, which is consistent with the computed nucleation barriers.

In the second row of Fig.~\ref{fig:hs}, we discuss our results with respect to the predictions of CNT. Within CNT assuming spherical droplets, one finds
\begin{gather}
  k = \rho_lZ_cD_ce^{-\beta\Delta F_c}, \qquad
  \Delta F_c = \frac{16\pi\gamma}{3(\Delta P)^2}, \\
  Z_c = \sqrt{\frac{\Delta P}{6\pi \rho_s n_c}}, \qquad
  D_c = \frac{24D_l}{\lambda^2} n_c^{2/3}
\end{gather}
for the nucleation rate $k$, critical nucleation barrier $\Delta F_c$, the Zeldovich factor $Z_c$, and the diffusion coefficient $D_c=D(n_c)$ on top of the barrier. The latter is controlled by a thin layer of liquid particles surrounding the droplet, which can diffuse and ``lock'' themselves with a given rate onto the nucleus surface. For a spherical droplet, the number of liquid particles in this thin layer is $n_l=6n_c^{2/3}$ and the mean attachment time is $\tau=\lambda^2/(4D_l)$ with $D_l$ the self-diffusion coefficient of a liquid particle and $\lambda$ the characteristic length scale for a particle to bind to a crystal site. Additionally, the solid-liquid interfacial tension $\gamma_{\infty}$ for a planar interface is taken from Ref.~\cite{schmitz2015ensemble} as the mean value between the three different facets $100$ ($0.581k_BT$), $110$ ($0.559k_BT$), and $111$ ($0.544k_BT$). This leads to $\gamma_{\infty}\simeq0.56$. Finally, the self-diffusion coefficient $D_l$ is taken from Ref.~\cite{speedy1987diffusion} and the locking distance is fixed at $\lambda=0.28d$, which is close to reported values~\cite{espinosa2016seeding}.

In Figs.~\ref{fig:hs}(e-h), we plot the critical nucleus size $n_c$, the Zeldovich factor $Z_c$, the critical barrier $\Delta F_c$, and the diffusion coefficient $D_c$ extracted from our simulations. These quantities are plotted \emph{vs.} the pressure difference $\Delta P$ to test the principal scaling predictions of CNT. In all cases, we find our data to be consistent with the predicted scaling. We also note that all data are in agreement with values reported in the literature with some small discrepancies for $n_c$, $Z_c$, and $D_c$. These deviations are expected since these values are sensitive to the specific choice of parameters in the bond-order analysis (cf. Sec.~\ref{sec:cv}). Additionally, the values reported previously for $D_c$ are based on computing the mean-square displacement of $n$ on top of the free energy barrier, where an additional time averaging of the bond order is needed Ref.~\cite{dijkstra11}. As shown in Ref.~\cite{dijkstra10}, the nucleation barrier $\Delta F_c$ does not suffer from this ambiguity and overall we observe in Fig.~\ref{fig:hs}(g) an excellent agreement. Additionally, we believe that our extracted values for $D_c$ from the MFPT method are consistent since we observe a crossover, where for large droplets the diffusion is dominated by the nucleus surface (following $D_c\sim n_c^{2/3}\sim \Delta P^{-2}$) and for small droplets (large pressure difference), $D_c$ decreases more strongly since the relaxation time of the liquid starts to grow approaching glassy dynamics for $\phi>\phi_m$. However, even though the scaling of all quantities are consistent with CNT, we observe that the use of the capillary approximation employing $\gamma_{\infty}$ fails to accurately model the numerical data. Agreement is only reached when using the much larger, effective interfacial tension $\gamma_{\text{CNT}}\simeq0.76$. In fact, this effective interfacial tension is not constant and decreases at low packing fractions~\cite{espinosa2016seeding}. This discrepancy is the point of departure for the second part of this series.

% ---- Conclusion ----
\section{Conclusions and Outlook}

To conclude, we have discussed and compared two different strategies to reconstruct the free energy landscape of a one dimensional activated process through (i)~calculating the mean first passage time and (ii)~calculating the splitting probability. We have tested both methods for direct brute force and rare-event simulations employing forward flux sampling, and found excellent agreement between all of them. Specifically, we have demonstrated the validity of both approaches for a toy model, a single trapped particle moving in a double-well potential. We have then applied both methods to study the crystallization of hard spheres, demonstrating that the free energy barriers thus extracted are consistent with the ones computed from umbrella sampling. The second method has the computational advantage that both kinetic information (the crystallization rate) and static information (the critical nucleus size and the nucleation barrier) can be extracted from the same ensemble of non-equilibrium \emph{forward} trajectories. Quantities at the nucleation barrier follow the scaling predicted by classical nucleation theory although quantitative agreement requires to plug in an effective interfacial tension that is much larger than the bulk interfacial tension. Although we have focused on hard spheres as a model system, the method described here to extract static informations from biased rare-event simulations is applicable more widely and should be valuable in investigations of more complex systems.

In part \RNum{2}, we will further explore the discrepancy between effective and bulk interfacial tension and its role in the nucleation process. In particular, we will exploit the method presented here to gain access to the equilibrium properties of small to intermediate droplets, which, so far, have not been accessible. To this end we combine the nucleation barriers extracted in this first part with the nucleation theorem to evaluate the nucleation work for a much larger range of supersaturation. In our approach, we make use of, and combine, a variety of state-of-the-art numerical methods used in nucleation theory: FFS, the seeding method, and simulations of equilibrium droplets in finite systems. We finally discuss the behavior of the surface of tension for the crystallization of hard spheres, drawing an analogy from earlier studies for the gas-liquid nucleation.

%% ---- acknowledgments ----

\begin{acknowledgments}
We thank K. Binder for illuminating discussions and critical remarks. We gratefully acknowledge ZDV Mainz for computing time on the MOGON supercomputer. We acknowledge financial support by the DFG through the collaborative research center TRR 146.
\end{acknowledgments}

%% ---- appendix ----

\appendix

\section{Relation between $P^+$ and $P$}
\label{ap:pst_peq}

The dynamics of the population $P^+_q$ on a discrete one-dimensional chain between two end nodes $q_a$ and $q_b$ is governed by the master equation (birth-death process)
\begin{equation}
\frac{\partial P^+_q(t)}{\partial t}=j_{q_{i-1}}(t)-j_q(t),
\end{equation}
where $q$ denotes the index of a node. The local current $j_q$ passing through this node is express via
\begin{equation}
j_q(t)=f_qP^+_q(t)-g_{q+1}P^+_{q+1}(t),
\end{equation}
where $f_q$ are the forward attachment rates from node $q$, and $g_{q+1}$ the backward attachment rates from the node $q_{q+1}$. These rates are connected in equilibrium ($j_q=j=0$) via the detailed-balance condition $f_qP_q=g_{q+1}P_{q+1}$. In a steady state, the current is necessarily constant and can be expressed (inserting the detailed-balance condition) as
\begin{equation}
j=f_qP^+_q-f_{q}P_q\frac{P^+_{q+1}}{P_{q+1}}.
\end{equation}
We divide both sides by $f_qP_q$ and sum from $q<q_b$ to $q_b-1$ (where $q_b$ is the absorbing end node with $P^+_{q_b}=0$) to obtain
\begin{equation}
j=\frac{P^+_q}{P_q}\Bigg[\sum_{k=q}^{q_b-1}\frac{1}{f_kP_k}\Bigg]^{-1}.
\end{equation}
In the presence of a large barrier with rare crossings, one assumes $P_{q_a}\simeq P^+_{q_a}$, which allows to eliminate the current $j$ yielding the relation 
\begin{equation}
P^+_q=P_q\sum_{k=q}^{q_b-1}\frac{1}{f_kP_k}\Bigg/\sum_{k=q_a}^{q_b-1}\frac{1}{f_kP_k}
\end{equation}
linking the forward probability distribution $P^+_q$ to the equilibrium probability distribution $P_q$ for any node $q$. Replacing the index $q$ by a continuous variable and the sums by integrals then yields Eq.~(\ref{eq:pst_peq}).

\section{Splitting probabilities}
\label{ap:pa_pb}

Here we derive the return probability $P_A(q)$ to fall back directly to the basin $A$ from a node $q$, which will then gives us the absorption probability $P_B(q)$ through the condition $P_A+P_B=1$ at any $q$. We define the forward probability to jump from the node $q$ to $q+1$ by $u_q=f_q/(f_q+g_q)$ and the backward probability to fall back from the node $q$ to $q-1$ by $v_q=g_q/(f_q+g_q)$. The return probability $P_A(q)$ at the node $q$ is the solution of the recursive relation
\begin{equation}
  P_A(q)=u_qP_A(q+1)+v_qP_A(q-1)
\end{equation}
with boundary conditions $P_A(q_a)=1$ and $u_{q_a}=1$. One can identify the following relation,
\begin{equation}
  P_A(q+1)-P_A(q)=\frac{v_q}{u_q}[P_A(q)-P_A(q-1)],
\end{equation}
where a simple iteration leads to
\begin{equation}
  P_A(q+1)-P_A(q)=[P_A(q_a+1)-P_A(q_a)]\prod_{j=1}^{q}\frac{v_q}{u_q}.
\end{equation}
Summing this expression from node $q_a+1$ to an arbitrary node $>q$, we have
\begin{equation}
  \label{eq:pa1}
  P_A(q+1)-P_A(q_a+1)=[P_A(q_a+1)-1]\sum_{k=1}^{q}\prod_{j=1}^{k}\frac{v_j}{u_j}.
\end{equation}
We can extract a solution for $P_A(q_a+1)$ by noting that for large $q\to\infty$ or absorbing boundary $q=q_b$ the probability to fall back to $A$ is zero, $P_A(\infty)=0$, which gives
\begin{equation}
\label{eq:pa2}
P_A(q_a+1)=\frac{\sum_{k=1}^{\infty}\prod_{j=1}^{k}\frac{v_j}{u_j}}{1+\sum_{k=1}^{\infty}\prod_{j=1}^{k}\frac{v_j}{u_j}}.
\end{equation}
By combining Eqs.~\ref{eq:pa1} and~\ref{eq:pa2} and rewriting $v_q/u_q$ as $g_q/f_q$, we derive the general expression for any $q$,
\begin{equation}
P_A(q)=\frac{\sum_{k=q}^{\infty}\prod_{j=1}^{k}\frac{g_j}{f_j}}{1+\sum_{k=1}^{\infty}\prod_{j=1}^{k}\frac{g_j}{f_j}}.
\end{equation}
Finally, we employ the detailed-balance condition to rewrite $\prod_{j=1}^{k}\frac{g_j}{f_j}$ in order to obtain
\begin{equation}
P_A(q)=\sum_{k=q}^{\infty}\frac{1}{f_kP_k}\Bigg/\sum_{k=q_a}^{\infty}\frac{1}{f_kP_k}=\frac{P^+(q)}{P(q)}.
\end{equation}

\section{Free energy reconstruction}
\label{ap:fe_reconstruct}

\begin{figure}[b!]
  \includegraphics[scale=1.0]{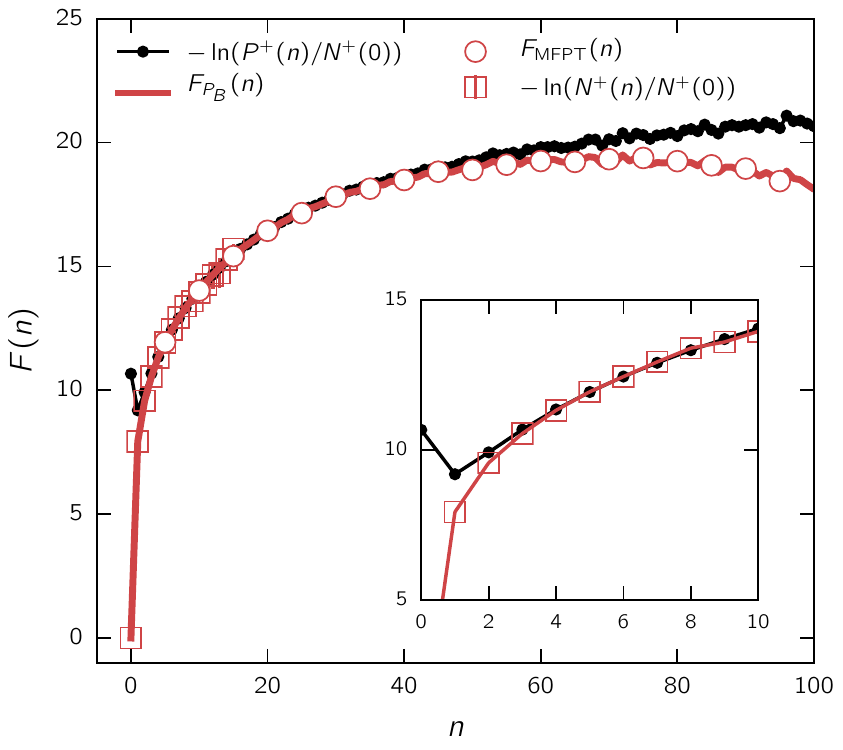}
  \caption{\textbf{Free energy reconstruction.}~Evolution of $-\ln[N^+(n)/N^+(0)]$, $-\ln[P^+(n)/N^+(0)]$ and the two free energy reconstructions $F_{P_B}(n)$ and $F_{\text{MFPT}}(n)$ as function of the nucleus size $n$. The inset show in more detail the discrepancy between $-\ln[N^+(n)/N^+(0)]$ and  $-\ln[P^+(n)/N^+(0)]$ for small nuclei.}
  \label{fig:fe_reconstruct}
\end{figure}

The free energy barrier for the formation of a cluster of size $n$ is connected to $N(n)$, the average number of clusters of size $n$, via $F(n)=-\ln[N(n)/N(0)]$~\cite{reiss1999some,auer2004numerical}. It differs at small cluster sizes from the free energy that we extract by projecting the transition path onto the largest nucleus $F_{\text{big}}(n)$. The reason is that, first, there can be more than one small cluster of size $n$ in one configuration and, second, it is likely that small clusters are not detected since there is always a larger cluster detected as the largest one. This makes $F_{\text{big}}(n)$ being biased towards larger values at small $n$, see inset of Fig.~\ref{fig:fe_reconstruct}. This effect will be even stronger for larger systems, which shifts the average size of the largest cluster to larger values~\cite{maibaum2008comment}. For sufficiently large cluster sizes, the probability to observe an even bigger cluster is negligible and both free energies coincide. To reconstruct the free energy $F(n)$ from the measured $F_{\text{big}}(n)$ for the full range of cluster sizes, we follow Ref.~\cite{lundrigan2009test,leitold2016nucleation}. We first compute $N^+(n)$, the steady state average number of clusters of size $n$, through
\begin{equation}
  N^+(0) + \sum_{i=1}^{n_b} iN^+(n_i) = N.
\end{equation}
We then patch together $-\ln[N^+(n)/N(0)]$ for small clusters, where we know that $N^+(n)\simeq N(n)$, to $-\ln[P(n)/N(0)]$ for larger clusters,
\begin{equation}
  F(n) = \begin{cases}
    -\ln[N^+(n)/N(0)] & n<c \\
    -\ln[P(n)/N(0)] & n\ge c,
  \end{cases}
\end{equation}
where $c$ is the cluster size at which $P(c)\simeq N(c)$ which depends on the system size, bond order parameter, and state point. We show the typical result of this method in Fig.~\ref{fig:fe_reconstruct}, and additional information can be found in Refs.~\cite{lundrigan2009test,leitold2016nucleation}.

%% ---- bibliography ----

\end{document}